\input{aipcheck}

\documentclass[
    ,final            
  ]
  {aipproc}

\layoutstyle{6x9}

\begin{document}

\title{Coherent J/$\psi$ photoproduction in ultra-peripheral Pb-Pb collisions\\ at $\sqrt{s_{\mathrm{NN}}}$ = 2.76 TeV~\footnotemark\footnotetext{Presented at DIFFRACTION 2012: International Workshop on Diffraction in High-Energy Physics. Puerto del Carmen, Canary Islands, 10-15 September 2012.}}

\classification{14.40.Pq, 13.40.-f,13.25.Gv,25.75.-q}
\keywords      {Ultra-peripheral collisions, heavy quarkonia}

\author{J.D. Tapia Takaki (for the ALICE Collaboration)~\footnotemark~\footnotetext{Daniel.Tapia.Takaki@cern.ch}}{
  address={Institute de Physique Nucl\'eaire d'Orsay (IPNO), \\ Universit\'e Paris-Sud, CNRS-IN2P3, Orsay, France.}
}

\begin{abstract} The first LHC measurement on ultra-peripheral heavy-ion collisions was carried out with the ALICE experiment. In this paper, ALICE results 
on exclusive J/$\psi$ studies in Pb-Pb collisions at $\sqrt{s_{\mathrm{NN}}}$ = 2.76 TeV, in the rapidity region -3.6 $<$ $y$ $<$ -2.6, are given. 
The coherent J/$\psi$ cross section was found to be d$\sigma^{coh}_{J/\psi}/dy$ = 1.00 $\pm$ 0.18 (stat) $^{+0.24}_{-0.26}$ (syst) mb. 
These studies favour theoretical models that include strong modifications to the nuclear gluon density, also known as nuclear gluon shadowing. 
\end{abstract}

\maketitle


\section{Introduction}

Ultra-peripheral heavy-ion collisions (UPC) can take place when the ions pass by each other with impact parameters larger than the sum 
of their radii. UPC reactions are governed by two-photon and photonuclear interactions, while those of hadronic nature are strongly suppressed~\cite{Review2005,Review2008}.
This is so as the electromagnetic field of the nucleus enhances the intensity of the virtual photon flux. Note that the number of
photons scales like $Z^{2}$, where $Z$ is the nucleus charge. The virtuality of the photon is 1/$R$ $\sim$ 30 MeV/$c$, where R is 
the radius of the nucleus.

In recent years, there has been an increasing interest in ultra-peripheral collisions for various colliding systems at different energies. 
In particular, exclusive vector meson production in heavy-ion reactions is expected to probe the nuclear gluon density~\cite{Ryskin:1992ui,Martin:2007sb}, 
for which their is considerable uncertainty at low values of Bjorken-$x$. At the forward rapidities studied here, the relevant values of $x$ are 
$\sim$10$^{-2}$ and $\sim$10$^{-5}$. Note that either nucleus can serve as photon emitter or photon target. Exclusive J/$\psi$ production 
has been recently measured in Au-Au collisions at RHIC~\cite{Afanasiev:2009hy}. However, their measurement suffered from very low statistics. Hence, no 
conclusions concerning nuclear shadowing were made. 

This paper will focus on the first LHC measurement on exclusive photoproduction of J/$\psi$ vector mesons produced in Pb-Pb collisions 
at $\sqrt{s_{\mathrm{NN}}}$ = 2.76 TeV~\cite{Abelev:2012ba}. This paper begins by describing the experimental apparatus and the collected data, it will then go on to discuss the analysis steps and the way the coherent J/$\psi$ cross section was obtained. This paper will also mention how J/$\psi$ photoproduction is treated by several theoretical models, as well as giving the comparison of the measured cross section to the available predictions. Finally, a summary will be given. 

\section{ALICE detector}

The ALICE detector is described elsewhere~\cite{Aamodt:2008zz}. The ALICE muon system consists of a spectrometer covering the pseudo-rapidity 
region -4.0 $< \eta <$ -2.5. It was designed to identify muons with a momentum larger than 4 GeV/$c$. It consists of a front absorber followed by a 3 T$\cdot$m dipole magnet, five tracking stations based on Cathode Pad Chambers, a passive muon-filter wall, and two trigger stations composed of Resistive Plate Chambers. In this analysis, the Silicon Pixel Detector,
the Zero Degree Calorimeters and the VZERO counters were also used. The VZERO consists of two arrays of 32 scintillator counters each, which 
are placed around the beam pipe on either side of the interaction region: VZERO-A and VZERO-C cover the pseudo-rapidity range 2.8 $< \eta <$5.1 and
-3.7$< \eta <$-1.7, respectively. Finally, two sets of hadronic Zero-Degree Calorimeters (ZDCs) are located at 116 m on either side of the Interaction Point. 
These detect neutrons emitted in the very forward region, for example neutrons emitted following electromagnetic dissociation.

\section{Analysis steps and results}

\begin{figure}
  \includegraphics[height=.3\textheight]{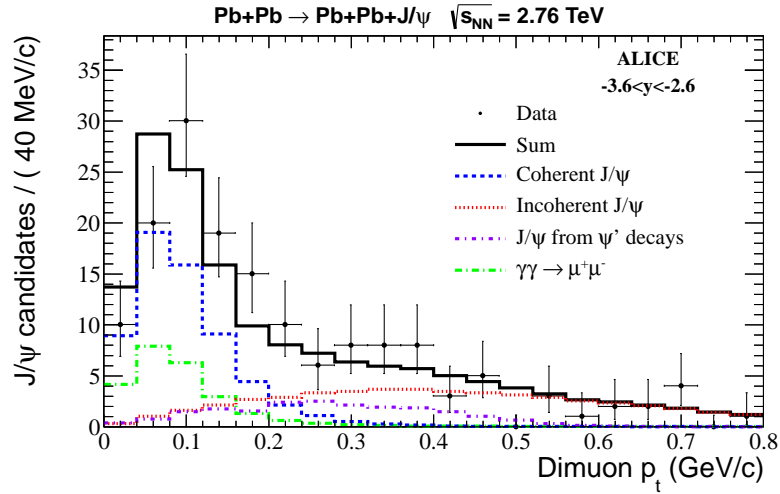}
   \label{fig:Figure2}
  \caption{Dimuon $p_\mathrm{T}$~distribution for events satisfying the event selection described in the text,
with the $p_\mathrm{T}$-range extended to $p_{\rm T} <$~0.8~GeV/$c$. The data points are fitted using
four different Monte Carlo templates: coherent J/$\psi$~production (dashed - blue), incoherent J/$\psi$~production
(dotted - red), {J/$\psi$}s~from $\psi^{'}$ decay (dash-dotted - violet), and $\gamma \gamma \rightarrow \mu^+ \mu^-$ 
(dash-dotted - green). The solid histogram (black) is the sum. The fitting function shapes were provided by STARLIGHT events~\cite{starlightMC} folded with realistic simulations.}
\end{figure}

This analysis was carried out on a data sample corresponding to an integrated luminosity of 55 $\mu$b$^{-1}$ collected during the 2011 Pb running. A dedicated UPC trigger was active to select dimuons in an otherwise empty detector, both from $\gamma \gamma$ and from J/$\psi$~decays. It demanded the following event characteristics: (i) a single muon trigger above a 1 GeV/$c$ $p_{\mathrm{T}}$-threshold; (ii) at least one hit in the VZERO-C detector, and (iii) no hits in VZERO-A. A detailed description of the event selection can be found in~\cite{Abelev:2012ba}. Standard cuts to ensure a good muon selection were required~\cite{Aamodt:2011gj}, followed by other cuts to suppress the remaining hadronic background. In particular, the VZERO offline timing was required to be compatible with crossing beams, and only events with a neutron ZDC signal below 6 TeV were kept. This cut does not remove any events with a J/$\psi$ produced with a transverse momentum below 0.3 GeV/$c$, but reduces hadronic contamination at higher $p_{\mathrm{T}}$. This analysis deals with exclusive reactions, \textit{i.e.} Pb+Pb $\rightarrow$ Pb+Pb + J/$\psi$. For this reason, only events with exactly two oppositely charged muons were selected. After applying these selections, the J/$\psi$ yield was described as the sum of four physics processes: coherent and incoherent J/$\psi$ photoproduction, $\gamma \gamma$ continuum, and J/$\psi$ candidates from the $\psi^{'}$ decay. Coherent J/$\psi$ candidates are those where the photon couples coherently to all nucleons in the reaction (target nucleus normally does not break up). On the other hand, incoherent J/$\psi$ photoproduction occurs when the photon couples to a single nucleon, $\textit{i.e.}$ the quasi-elastic scattering off a single nucleon. In this case, the target nucleus normally breaks up, but except from single nucleons or nuclear fragments in the very forward region no other particles are produced. In both cases the J/$\psi$ $p_{\mathrm{T}}$ is very low; the expected average transverse momentum is about 60~MeV/$c$ and 500~MeV/$c$ for coherent and incoherent photoproduction, respectively. Figure~\ref{fig:Figure2} shows the dimuon $p_{\mathrm{T}}$~distribution integrated over 2.8 $<$ $M_{\mathrm{inv}}$~$<$ 3.4~GeV/$c^{2}$. The clear peak at low $p_{\mathrm{T}}$~is due to coherent interactions, while the tail extending out to 0.8 GeV/$c$ comes from incoherent production. Taking also into account the finite detector resolution, dimuons where required to have $p_{\mathrm{T}}<$~0.3~GeV/$c$, resulting in 117 J/$\psi$ candidates. By performing a fit to the corresponding dimuon invariant mass spectrum, the extracted number of J/$\psi$ is $N_{\mathrm{yield}}$ = 96 $\pm$ 12 (stat) $\pm$ 6 (syst). To perform the fit, a maximum likelihood method on the set of events satisfying the selections described was used, considering a Crystal Ball function to describe the signal and an exponential to account for the underlying continuum. The exponential slope parameter of the $\gamma \gamma$ continuum is -1.4~$\pm$~0.2~GeV$^{-1} c^2$ in good agreement with the corresponding Monte Carlo expectation (-1.39~$\pm$~0.01~GeV$^{-1} c^2$). This indicates the absence of hadronic background in the kinematic regions considered. The estimated feed-down contribution from the $\psi^{'}$ decay is 11 $\pm$ 6\% for dimuon $p_{\mathrm{T}}<$~0.3~GeV/$c$, while the corresponding incoherent contribution is 12 $^{+14}_{-4}$\%. 
 
The QED continuum pair production was used for normalisation purposes. Although in principle it can be calculated with high accuracy, the fact that the photon coupling
to the nuclei is Z$\sqrt{\alpha}$ (with Z =82 here) rather than just $\sqrt{\alpha}$ increases the uncertainty from higher order terms. There is also an
uncertainty associated with the minimum momentum transfer and the nuclear form factor. For these reasons, the uncertainty in the STARLIGHT two-photon cross section
was estimated to be 20\%~\cite{Abelev:2012ba}. The other sources of systematic errors are given in~\cite{Abelev:2012ba}. The differential coherent J/$\psi$ cross section is $\mathrm{d}\sigma_{J/\psi}^{\mathrm{coh}} /\mathrm{d}y = 1.00 \pm 0.16(\mathrm{stat}) ^{+0.24}_{-0.26}(\mathrm{syst})$~mb. 

\begin{figure}
  \includegraphics[height=.3\textheight]{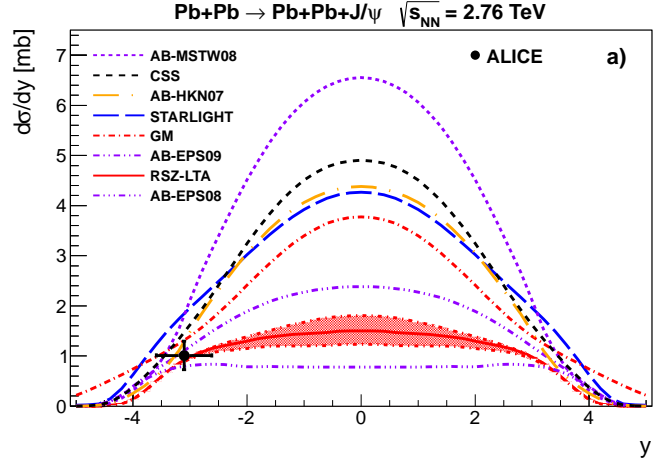}
  \caption{Measured coherent differential cross section of J/$\psi$~photoproduction in
ultra-peripheral Pb-Pb collisions at $\sqrt{s_{\mathrm{NN}} } = 2.76$ TeV. The error is the quadratic sum of the
statistical and systematic errors. The theoretical calculations described in the text are also shown. The
rapidity distributions are shown.}
\label{fig:Figure3a}
\end{figure}

\begin{figure}
  \includegraphics[height=.3\textheight]{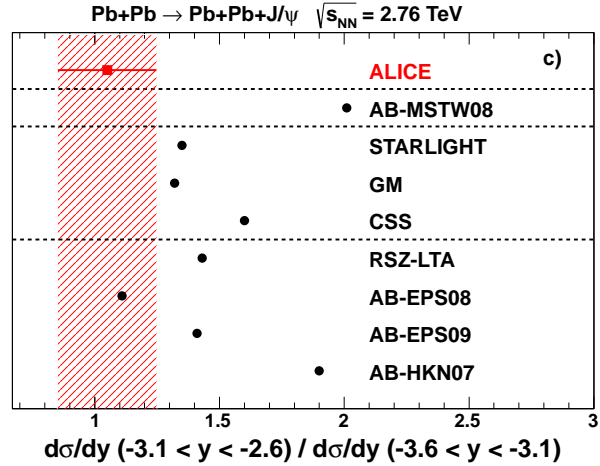}
  \caption{Ratio of the cross sections in the rapidity intervals -3.1 $< y <$ -2.6 and  -3.6 $< y <$ -3.1. The dashed lines in 
the lower two plots indicate the three model categories discussed in the text.}
\label{fig:Figure3c}
\end{figure}

Figure~\ref{fig:Figure3a} shows the comparison between the measured cross section and various models that calculate the photon spectrum in impact parameter space
in order to exclude interactions where the nuclei interact hadronically. They differ mainly by the way in which the photonuclear interaction is treated. They can be grouped into
three different categories: i) those that include no nuclear effects (AB-MSTW08)~\cite{Adeluyi:2012ph}. Here all nucleons contribute to the scattering, and the forward scattering differential cross section scales with the number of nucleons squared; ii) those that use a Glauber approach to calculate the number of nucleons contributing to the scattering (STARLIGHT~\cite{starlightMC,starlight}, GM~\cite{Goncalves:2011vf} and CSS~\cite{Cisek:2012yt}); and those from partonic models, where the cross section is proportional to the nuclear gluon distribution squared (AB-EPS08~\cite{Goncalves:2011vf}, AB-EPS09~\cite{Goncalves:2011vf}, AB-HKN07~\cite{Goncalves:2011vf} and RSZ-LTA~\cite{Rebyakova:2011vf}). As expected, the sensitivity to shadowing at forward rapidities is reduced compared to that at mid-rapidity. At forward rapidities there is a two-fold ambiguity in the photon energy and the momentum transfer from the nucleus acting as a photon target. The AB-MSTW08 and STARLIGHT predictions deviate by about three standard deviations, if the statistical and systematic errors are added in quadrature. The RSZ-LTA, AB-EPS08 and AB-EPS09 models show the best agreement to data within one standard deviation. In addition, Figure~\ref{fig:Figure3c} shows the ratio of the cross sections in two rapidity intervals, $R = \sigma(-3.1<y<-2.6)/ \sigma(-3.6<y<-3.1) = 1.36 \pm 0.36 (\mathrm{stat})$ $\pm$ 0.19 $(\mathrm{syst})$. Here, only AB-MSTW08 and AB-HKN07 deviate by more than one standard deviation. Taken together, these results indicate that the AB-MSTW08 model is strongly disfavoured, while the predicted STARLIGHT cross section deviates by nearly three standard deviations. Best agreement is found with models that include nuclear gluon shadowing, in agreement with EPS08 or EPS09 parameterisations~\cite{Eskola:2009uj}. The findings of this first UPC study also suggest that in the future other heavier particles like the $\Upsilon$(1S) or even the Higgs boson could be measured when using the LHC as a ``photon collider".

\section{Summary}

In summary, ALICE has provided the first LHC measurement on exclusive J/$\psi$ photoproduction in Pb-Pb collisions at $\sqrt{s_{\mathrm{NN}}}$ = 2.76 TeV. The results of this 
study indicate that theoretical models on J/$\psi$ in UPC that include strong modifications to the nuclear gluon density are favoured. A future study investigating coherent vector 
meson photoproduction at central rapidities would be very interesting for understanding these effects. 

\begin{theacknowledgments}
I would like to thank the organisers of Diffraction 2012 for the opportunity 
to speak on this novel LHC measurement carried out by the ALICE collaboration.
\end{theacknowledgments}

\bibliographystyle{aipproc}   

\end{document}